# Using the quantization error from Self-Organized Map (SOM) output for detecting critical variability in large bodies of image time series in less than a minute

Birgitta Dresp-Langley* and John Mwangi Wandeto+

*ICube Lab CNRS and University of Strasbourg, Strasbourg, France*
*+Dedan Kimathi University of Technology, Nyeri, Kenya*

**E-mail for correspondence:** birgitta.dresp@icube.unistra.fr

**Keywords:** Self-Organized Map (SOM); quantization error; image time series; spatial contrast; variability; change detection

**Abstract**

The quantization error (QE) from SOM applied on time series of spatial contrast images with variable relative amount of white and dark pixel contents, as in monochromatic medical images or satellite images, is proven a reliable indicator of potentially critical changes in images across time and image homogeneity. The QE is shown to increase linearly with the variability in spatial contrast contents of images across time when contrast intensity is kept constant across images.

**Introduction**

The self-organized map (SOM) provides an architecture for artificial neural networks, designed for simulation experiments and a great many practical applications (**Kohonen, 1990**). The SOM effectively creates spatially organized internal representations of various features of input signals (auditory, tactile, visual) and is well suited for the rapid hierarchical clustering or classification of data. The self-organization process is able to detect relationships between features in images and is well suited for simulating complex processes such as brain mapping, semantic mapping, and, in particular, competitive learning. The self-organizing map algorithm orders responses spatially through best matching cell selection and adaptation of the synaptic weights to the learnt input. The use of self-organized maps is well-suited for modeling dynamic principles of brain analysis for visual organization and processes of perceptual learning (e.g. Dresp & Grossberg, 1997; Dresp, 1997; Dresp, 1998a; Wehrhahn & Dresp, 1998; Dresp, 1998b; Dresp, 1999a; Dresp, 1999b; Dresp & Grossberg, 1999; Pins, Bonnet & Dresp, 1999; Dresp, 2000; Fischer & Dresp, 2000; Dresp & Fischer, 2001; Dresp & Spillmann, 2001; Bonnet & Dresp, 2001; Dresp; 2002; Tzvetanov & Dresp, 2002; Dresp, Durand & Grossberg, 2002; Dresp-Langley & Durup, 2009; Spillmann, Dresp-Langley & Tseng, 2015; Dresp-Langley, 2015; Dresp-Langley & Grossberg, 2016; Dresp-Langley, 2016; Dresp-Langley, Reeves & Grossberg, 2017). Self-organized maps can be implemented for a variety of neural architectures and the computational time necessary for obtaining SOM output data for a large number of images is astonishingly short (i.e. less than a minute for 20 images, for example).

In previous work (Wandeto et al., 2017a, Wandeto et al., 2017b), we have shown that the quantization error (QE) from SOM output after image learning can be effectively exploited for the fast and reliable detection of local changes in time series of medical images and satellite images for specific geographic regions of interest. The goal of the following proof-of-concept study is to show that the QE varies consistently, reliably, and predictably with local variations in spatially distributed contrast signals in random-dot images and images with regularly distributed spatial contrast (geometric configuration). On the grounds of these systematic variations, it will be shown that the QE is a highly sensitive and reliable indicator of local and global image homogeneity: as images from a time series become more heterogeneous in spatial contents, the QE in the SOM output after learning consistently increases; conversely, as images from a time series become more homogenous in spatial contents, the QE consistently decreases.

**Materials and methods**

All images generated for the study had the same size in terms of x, y screen pixel coordinates (792 x 777 pixels). Time series of images with variable amount of black and white pixel contents were generated. In a given time series, the percentage of spatially distributed white and black pixel contents was increased systematically, the contrast intensity was kept constant within and across image series.

*Randomly distributed spatial contrast with increasing white pixel contents*

In a first time series of six images (Fig 1), the percentage of randomly distributed white pixel content increases regularly from +10% in the second image to +60% in the last image of the series, starting from an original reference image, which is the first of that series.

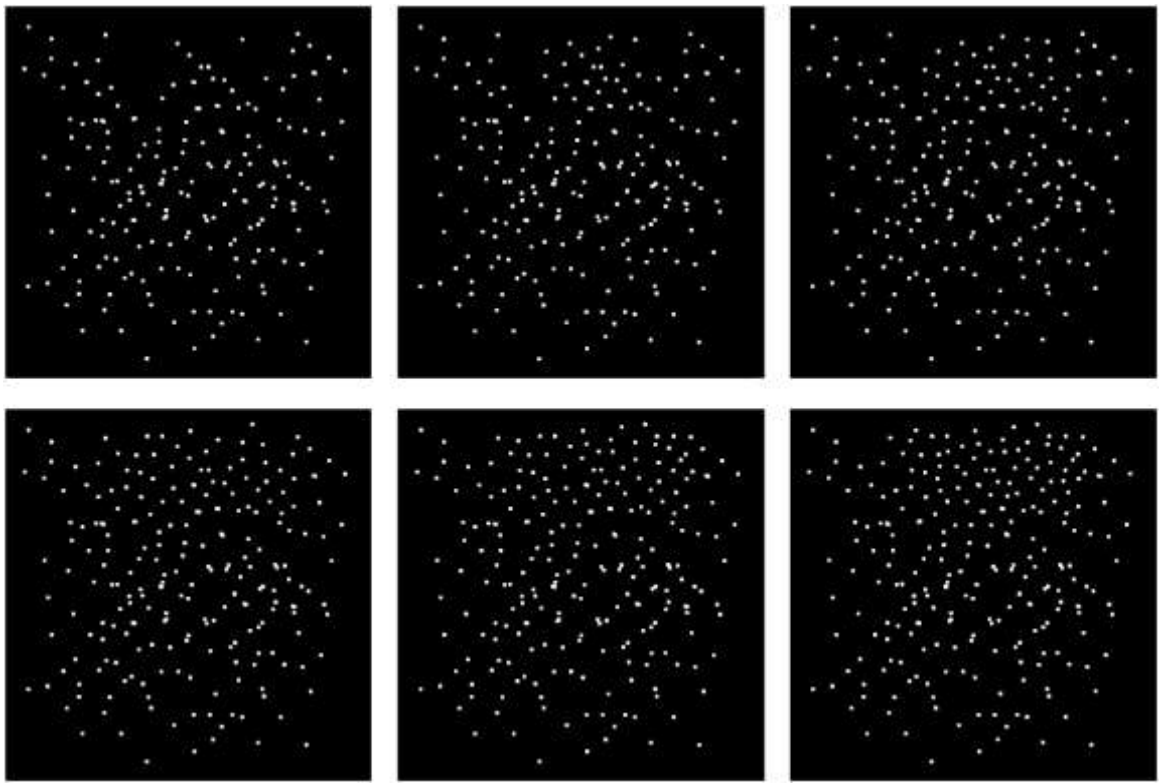

**Figure 1:** The randomly distributed white pixel content progressively increases between the first and last image of the series.

*Randomly distributed spatial contrast with increasing black pixel contents*

In the second time series, with three images (Fig 2), the percentage of randomly distributed black pixel content increases regularly from +20% in the second image to +30% in the third and last image of the series, starting from an original reference image, which is the first of the series.

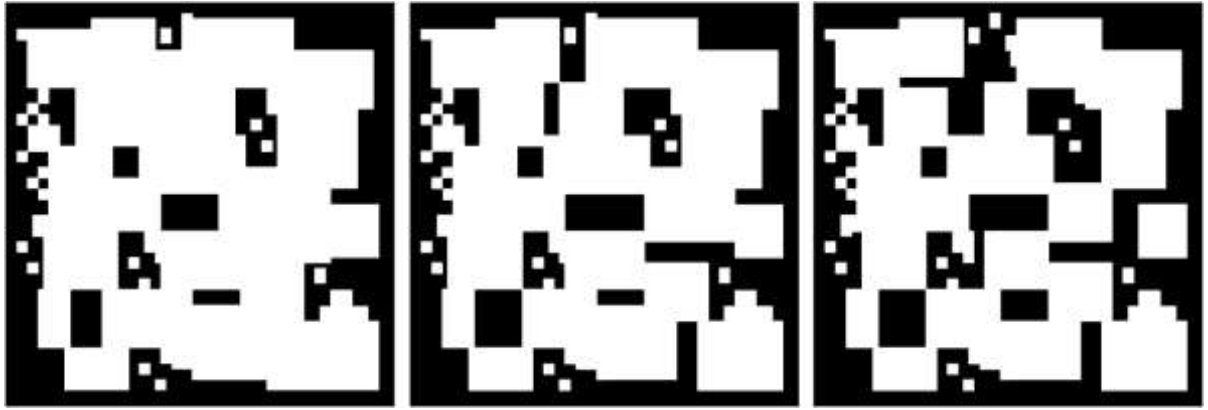

**Figure 2:** The randomly distributed black pixel content progressively increases between the first and last image of this series here.

*Spatial contrast pattern ("checkerboard") with increasing number of white squares*

In the third time series, with nine images (Fig 3), the percentage of systematically distributed white pixel content in the images increases regularly from 8% in the first image to 72% in the ninth and last image of the series.

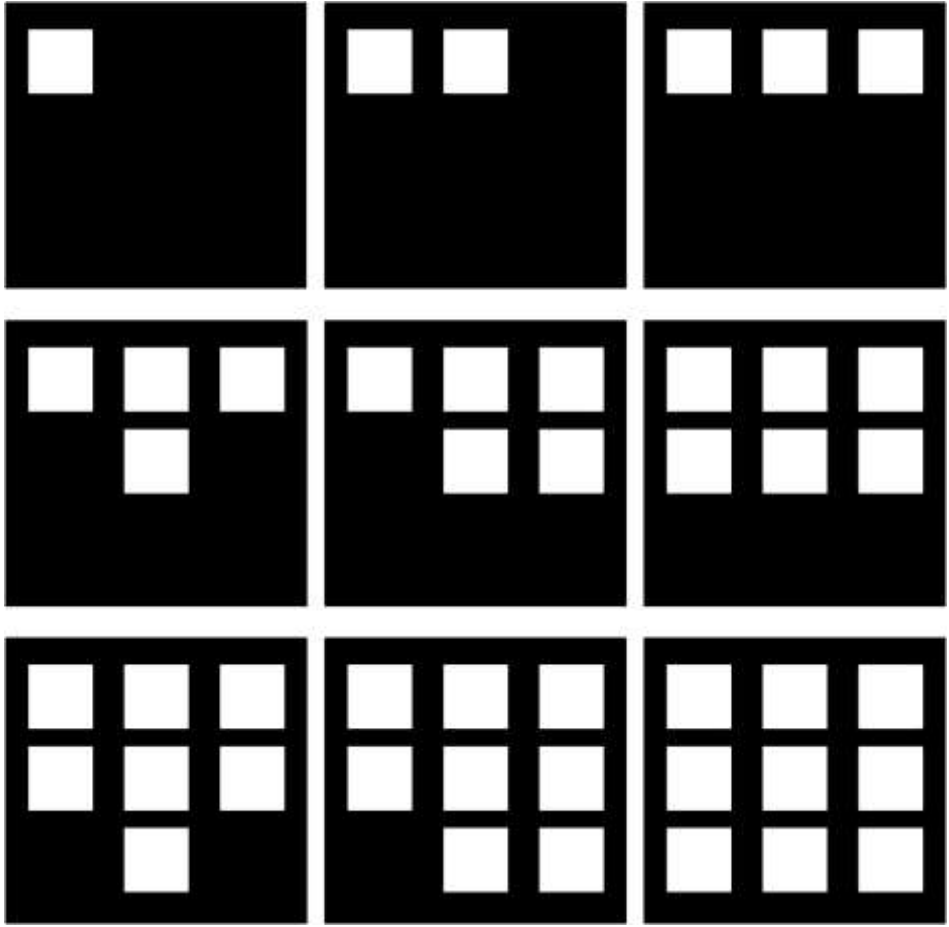

**Figure 3:** The percentage of systematically spaced white pixel content, producing the "checkerboard" patterns here, progressively increases between the first and last image of this series.

*Spatial contrast pattern ("checkerboard") with increasing size of white squares*

In the fourth time series, with nine images (Fig 4), the percentage of systematically spaced white pixel content in the images increases regularly from 2% in the first image to 18% in the ninth and last image of the series.

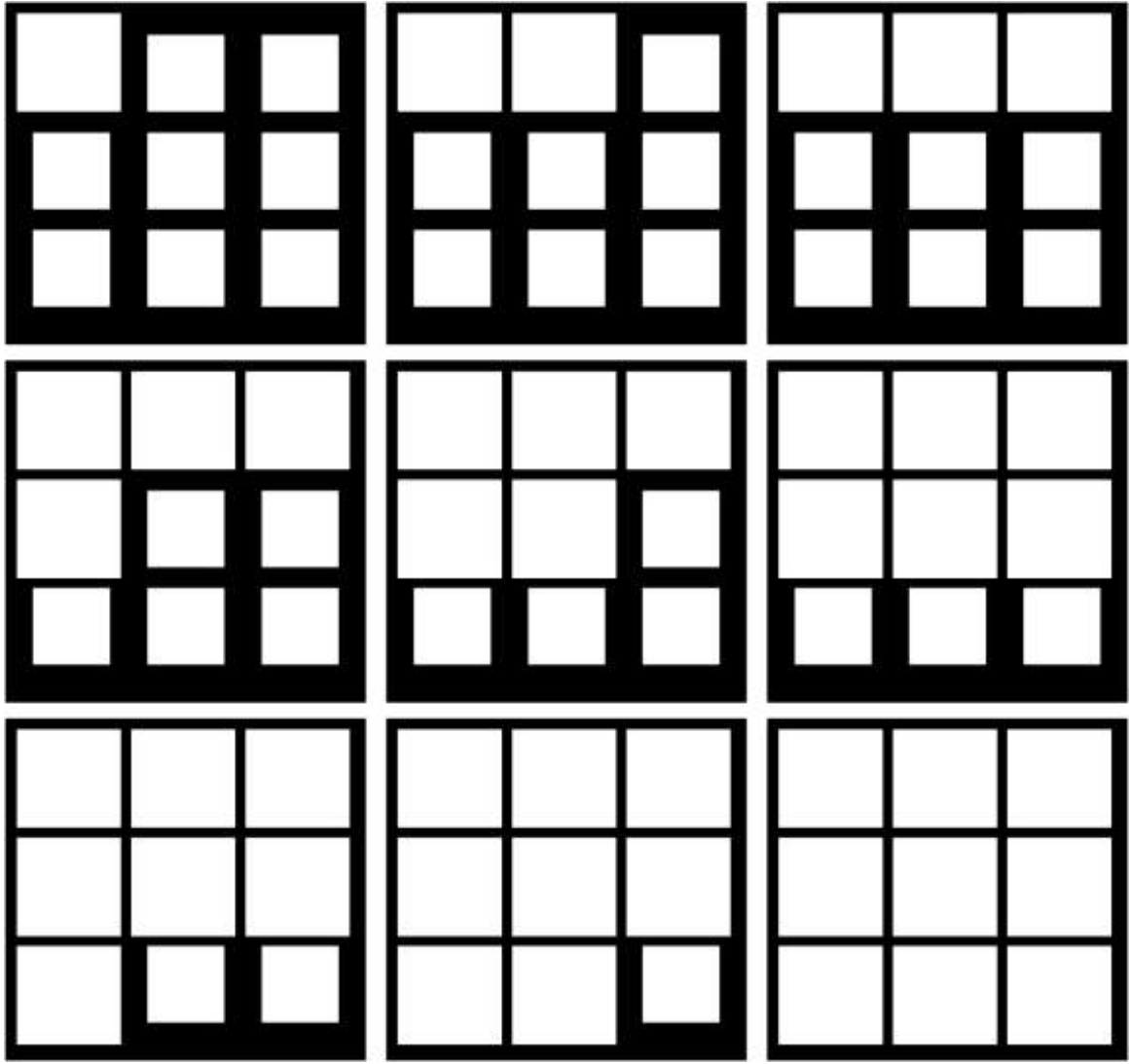

**Figure 4:** The percentage of systematically spaced white pixel content progressively increases between the first and last image, producing the systematic increase in single element size of the "checkerboard" patterns in this series here.

*Spatial contrast pattern with increasing size of a centrally placed white square*

In the fifth and last time series here, with six images (Fig 5), the percentage of centrally placed white pixel content in the images increases regularly from 1% in the first image to 32% in the sixth and last image of the series.

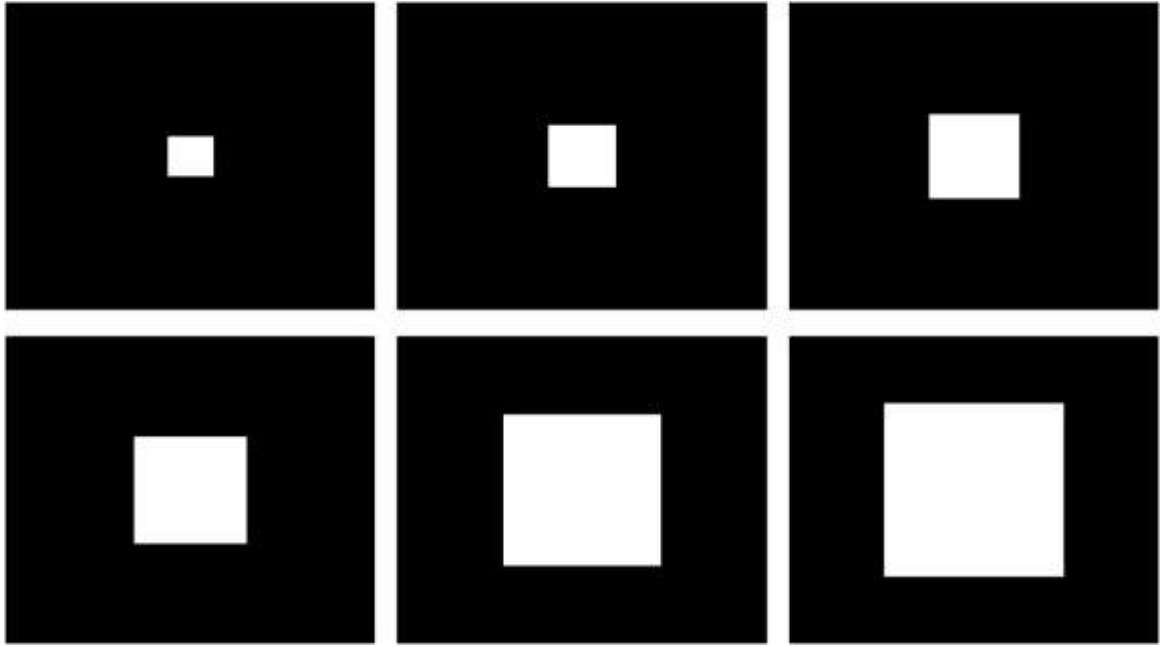

**Figure 5:** The percentage of the centrally placed white pixel content progressively increases between the first and last image in this series here.

*The Self-Organizing Map (SOM)*

A four-by-four SOM with sixteen artificial neurons was implemented, with an initial neighborhood radius of 1.2 and a learning rate of 0.2. The neural network was set to learn in 10.000 iterations, producing self-organized mapping output after learning in terms of final synaptic weights and the quantization error (QE). The results from these analyses, in terms of the quantization errors after learning obtained from SOM on images from each of the five different series, were represented graphically as a function of the image series SOM was run on.

**Results**

The quantization error from SOM on each image of each of the fie time series was plotted as a function of the increase (%) in white or black pixel content across a given series of images. These results are shown here below in Figures 6, 7, 8, 9 and 10.

It is shown that, as a given amount of spatially distributed pixel content increases acrss images of a given time series, the QE form the SOM output also increases. This is fully accounted for by a linear function with a near perfect goodness of fit with linear regression coefficients varying between 0.9598 and 0.996 (see Figures 6 - 10 here below).

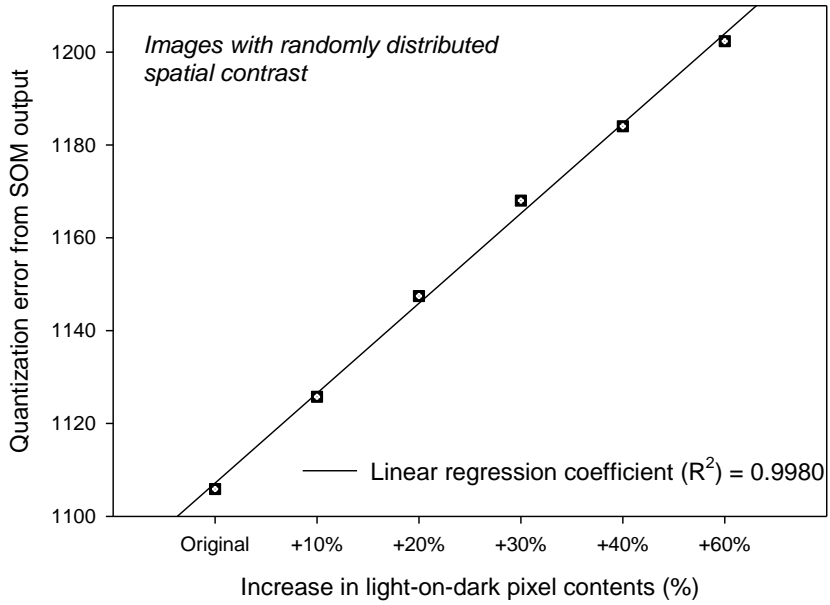

**Figure 6:** Results from SOM on the images of the first series. The QE increases linearly with increasing amount of white pixel content in the random contrast patterns.

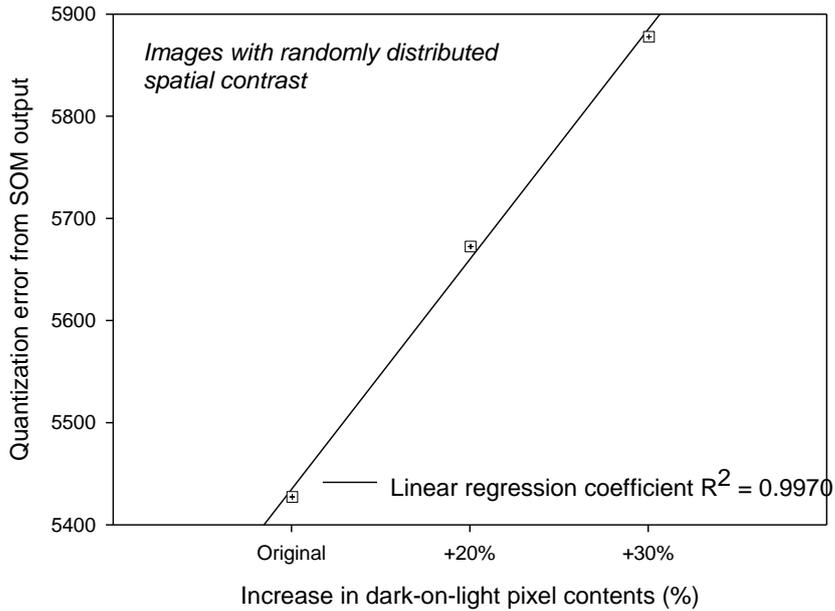

**Figure 7:** Results from SOM on the images of the second series. The QE increases linearly with increasing amount of black pixel content in the random contrast patterns.

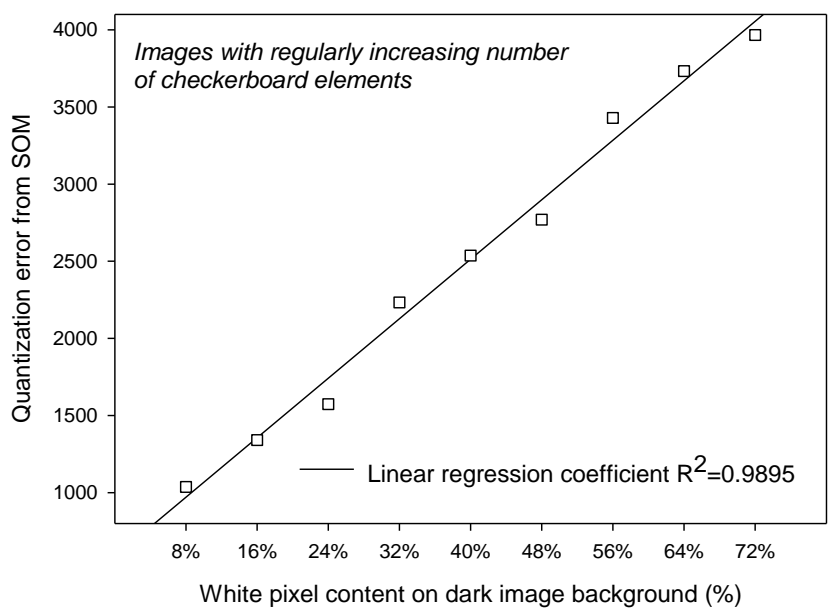

**Figure 8:** Results from SOM on the images of the third series. The QE increases linearly with increasing amount of white pixel content in the systematically spaced contrast patterns.

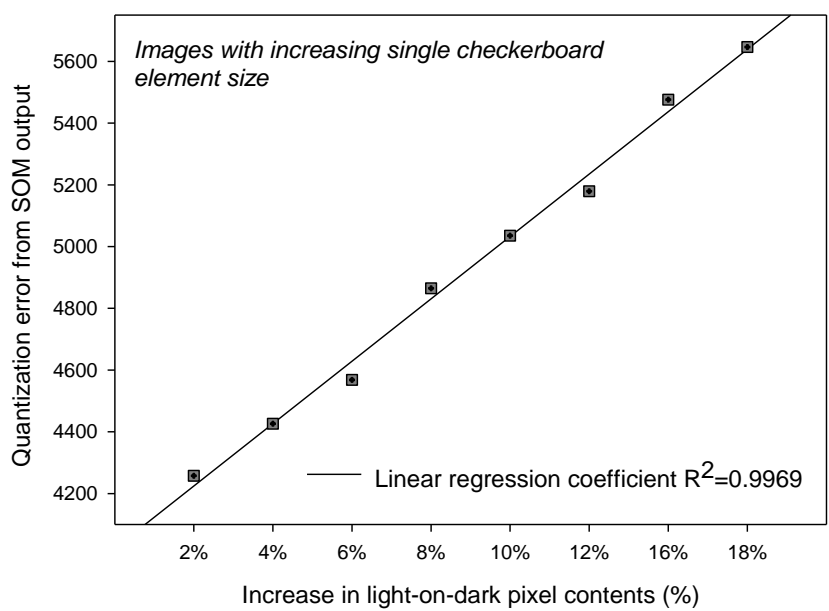

**Figure 9:** Results from SOM on the images of the fourth series. The QE, again, increases linearly with increasing amount of white pixel content in the spatial contrast patterns.

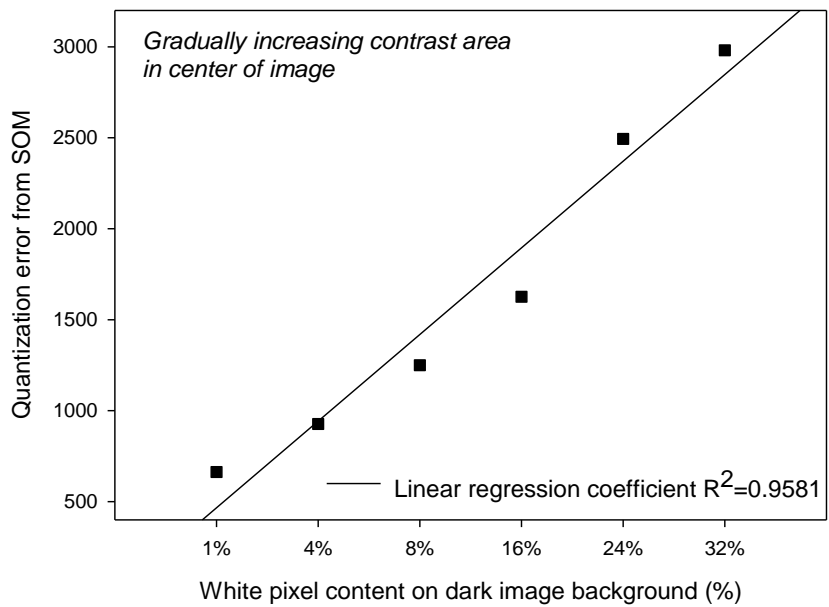

**Figure 10:** Results from SOM on the images of the fifth and last series. The QE increases linearly with increasing amount of white pixel content in the center of the images.

**Discussion**

The results of this proof-of-concept study clearly show that the quantization error (QE) from the SOM output is a highly reliable indicator of spatial contrast variability across images. When the contrast intensity is, as here, kept constant across images, the QE increase linearly as the variability in spatial contrast contents increases. These results point toward SOM as a fast, reliable and affordable preprocessing tool for large bodies of image data. This has important implications for society in all contexts where image data need to be interpreted swiftly and cost-effective with the goal of making potentially critical (medical geographic or other) data available fast to policy makers and the general public.